# Broadband resonances in ITO nanorod arrays

**Authors:** Shi-Qiang Li,[1,*] Kazuaki Sakoda,[2,3] John B. Ketterson,[3,4] Robert P. H. Chang[1,3,*1]

**Affiliations:**

[1]Department of Materials Science and Engineering, Northwestern University, 2220 Campus Dr., Evanston, IL, USA 60208-3108

[2]National Institute for Materials Science, 1-2-1 Sengen, Tsukuba, Ibaraki 305-0047, Japan

[3]NU-NIMS Materials Innovation Center, 2220 Campus Dr., Evanston, IL, USA, 60208-3108

[4]Department of Physics, Northwestern University, 2145 Sheridan Rd., Evanston, IL, USA 60208-3113

Abstract

In the nanophotonics community, there is an active discussion regarding the origin of the selective absorption/scattering of light by the resonances with nanorod arrays. Here we report a study of the resonances in ordered indium-tin-oxide (ITO) nanorod arrays resulted from the waveguide modes. We discover that with only 2.4% geometrical coverage, the micron-length nanorod arrays strongly interact with light across an extra-wide band from visible to mid-infrared resulting in less than 10% transmission. Simulations show excellent agreement with our experimental observation. Near-field profile obtained from simulation reveals the electric field is mainly localized on the surfaces of the nanorods at all the resonances. Theoretical analysis is then applied to explain the resonances and it was found that the resonances in the visible are different from those in the infrared. When the light arrives at the array, part of the light wave propagates through the free space in between the nanorods and part of the wave is guided inside the nanorods and the phase difference at the ends of the rod interactions forms the basis of the resonances in the visible region; while the resonances in the infrared are Fabry-Perot resonances of the surface guided waves between the two ends of the nanorods. The simple analytical formulae developed predict the spectral positions of these resonances well. This information can be used to design devices like wavelength-selective photodetector, modulators, and nanorod-based solar cells.

*Dr. Shi-Qiang Li: s-li@u.northwestern.edu, Prof. Robert Chang: r-chang@northwestern.edu

## Introduction

Low-dimensional materials are attracting significant interests in nanophotonics community,[1-5] due to their ability to guide and concentrate light in subwavelength length scale[6-8] and their size-dependent electronic and photonic properties.[9-11] In particular, one-dimensional materials like nanorods/nanowires are found to have a wide range of applications, including photovoltaic,[12, 13] electronics,[14] optoelectronics,[15] plasmonics,[16-18] and more.

In the last decade, many different nano-fabrication methods have been developed to rationally synthesize and assemble ordered and uniform nanorod arrays.[19-24]. One of the common issues discussed about the nanorod arrays is that many different processes are involved simultaneously in the light-array interactions, which often complicates the analysis. For example, not only does the detailed interaction depend on the shape of the nanorod and the arrangement of the nanorods, but it also depends on the material properties such as defect density, and electronic band gap.

Tin-doped indium oxide (ITO) is one of the best transparent conductive oxides. It is extensively used for photovoltaics and optoelectronics as electrode materials.[25-27] Recently, we demonstrated that it is possible to synthesize single-crystalline nanorod arrays made of ITO. They show superior plasmonic properties in the infrared[18] and tunable plasmonic-photonic coupling,[28] which can be used to engineer sharp plasmonic resonances.[29] In this study we focus on another interesting property of ITO nanorod arrays. We show that due to the permittivity dispersion, two different waveguide modes are supported by this system. They cause very similar spectral resonances but occurring at two very different optical regimes. In both cases, the fields are amplified on the surface of the nanorods at resonances. This implies potential applications for nanophotonic devices with active region around the surface of the nanorods, such as nano-photovoltaics with core-shell structure[8] and surface-enhanced plasmonic sensors.[30]

Experimental Results

ITO nanorod array with square lattice was used in this study. The lattice constant of array is 1.2 μm. The nanorods are 3 μm tall, with square cross-section (185 nm × 185 nm). The array is ordered and uniform as seen from the 30° tilted SEM image shown in Figure 1**(a)**. A zoomed-in view of the nanorod from the top is presented in Figure 1**(a)** (the left inset); it reveals the cross-section of the nanorod. The round object at the center of the nanorod is the catalyst – a gold nanoparticle used to grow the nanorod. (The nanorods were growth via Vapor-Liquid-Solid method on single crystalline yittra-stablized zirconia (YSZ) (100) substrate. The detailed growth conditions were reported previously.[18]) The second inset in the same figure is a high resolution TEM image showing the smooth sidewall of the nanorod, as well as the single crystallinity. The single crystalline nature of the nanorods is quite important to minimize detrimental surface scattering and grain boundary scattering.

A transmission spectrum was taken from the sample and plotted in Figure 1**(b)**. Strong oscillating spectral feature was observed across the whole spectrum range until 700 nm. The oscillation strength increases with wavelength. No spectral feature was observed from our control (200 nm ITO film on the same substrate) , shown as the blue curve in Figure 1(b).

Although the spectra in Figure 1**(b)** were measured with incident wavevector parallel to the nanorod axial axis (*i.e.* the long axis), transmittance was also measured with sample tilted from 0° to 25°. The oscillations were found stationary with respect to the azimuthal but shifts slightly with polar angle, which implies they are not from the photonic modes due to the 2-D lattice formed by the array. Furthermore, they are independent of polarization.

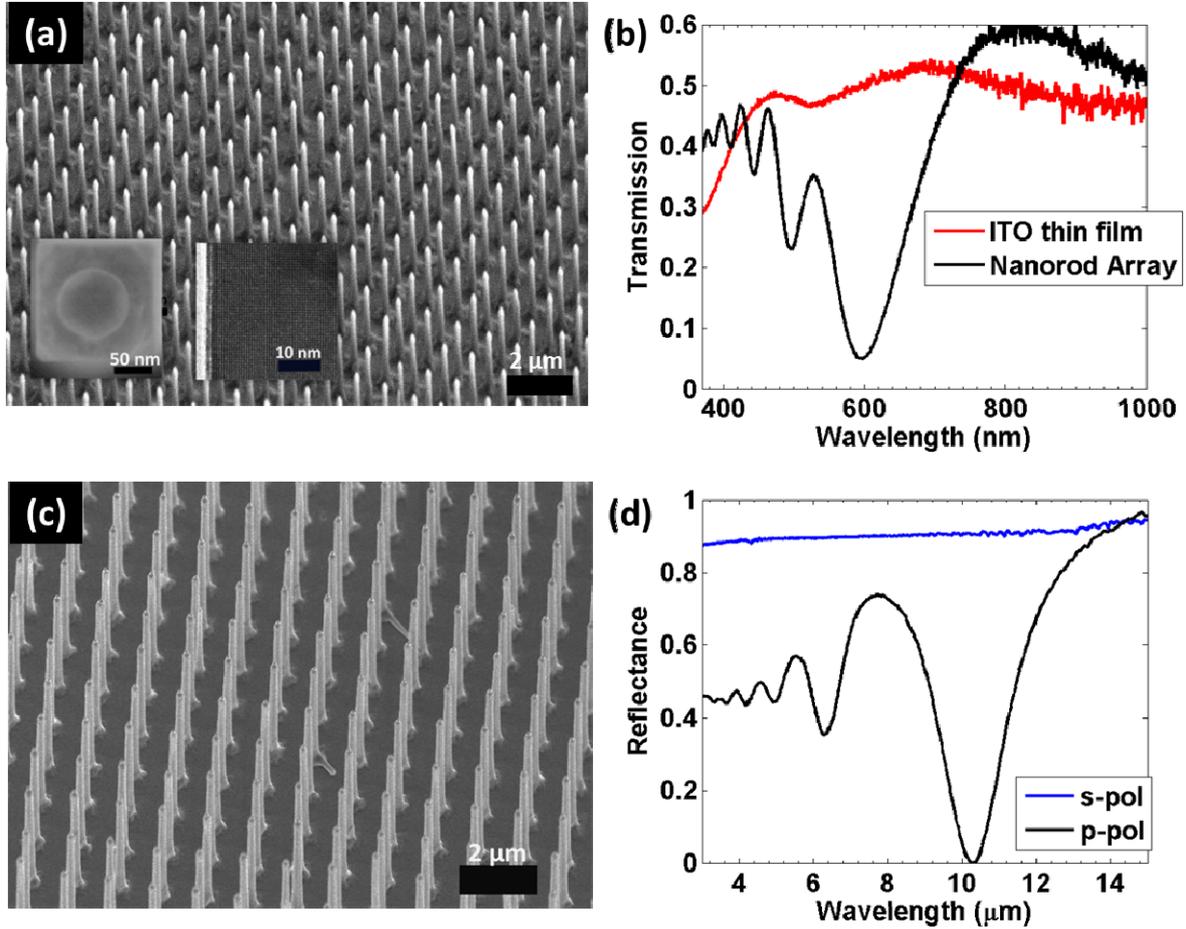

Figure 1. (a) SEM image of the first nanorod array studied here. (b) A transmission spectrum of the nanorod array, plotted together with an ITO film (200 nm) on the same substrate (YSZ). (c) SEM image of the second nanorod array studied here. (d) Specular reflectance spectra from the array shown in (c), with two different polarization.

The measurement of the reflectance of ITO arrays in the infrared showed similar resonances. We have chosen another array with longer nanorods (5 μm tall, Figure 1(**c**)) from which we observe a similar number of oscillations (Figure 1(**d**)). The measurement was done differently (by reflection) owing to the following reasons. Firstly, the substrate (YSZ) has strong phonon absorption beyond ~ 10 μm wavelength, so reflectance measurement is preferred. Secondly, since the ITO film is highly reflective in the mid-infrared due to the high density of free-carriers, a 200 nm thick ITO film was coated to enhance the reflectance. Thirdly, the incident wavevector is tilted 76° from the long axis of the nanorod as only TM (plasmonic) modes are supported. Fourthly, the resonance features in the infrared are polarization dependent. They are only excited when the electric field of the incident wave is polarized in the scattering plane (p-polarized, black curve in Figure 1(**d**)). Similar to those in the visible spectrum, the oscillation also gets stronger with a longer wavelength. Beyond 14 μm, no spectral feature was observed. The modes are not affected by the rotation of azimuthal angle.

## Mode Analysis and Simulation Results

The permittivity dispersion[18] of ITO is plotted in Figure 2(a), which is based on the following formula,

$$\varepsilon_{ITO}(\omega) = 3.95 - \frac{\omega_p^2}{\omega^2 + i\gamma\omega} \quad ,(1)$$

, where $\omega = \frac{2\pi c}{\lambda}$, $\omega_p$ is plasma frequency (1.52 eV), $\gamma$ is the damping factor (0.062 eV), $\lambda$ is the wavelength, $c$ is the speed of light in vacuum, and 3.95 is the high frequency dielectric constant.[31] We can see that the ITO in the visible spectral region is dielectric with small imaginary part of the permittivyt, and the real part of the permittivity decreases monotonically from 3.8 at 350 nm to around 3 at 700 nm. As the dimension (*d*) of nanorod cross-section is 185 nm, $\lambda/2d$ at this spectral range is equal or great than 1. With this small cross-section, only fundamental modes ($TE_{10}$ or $TM_{10}$) are supported in the nanorods. These modes are degenerated due to the symmetry of the cross-section.

Figure 2(**b**) shows the spectra calculated from the finite-element-simulation (FEM) simulation, which match very well with our experimental observation. The reflectance curve is also plotted

together, which has value smaller than 0.1. From the permittivity plot in Figure 2**(a)**, we note that trivial absorption will occur at this region, thus signal not collected by reflectance and transmittance mostly are scattered. This differs from previous studied semiconductor nanowire arrays, in which resonances leading to high absorption, and the resonances are attributed to whispering gallery modes[32] and Fabry-Perot modes.[33]

We also simulated how the modes behave with increased absorption coefficient, which we can set artificially in simulation. The results are shown in Figure 2(d). It was found that higher order modes are quite sensitive to the increased absorption coefficient and disappear quickly with the absorption coefficient $k$ increased to 0.1. The first order mode broadens and redshifts.

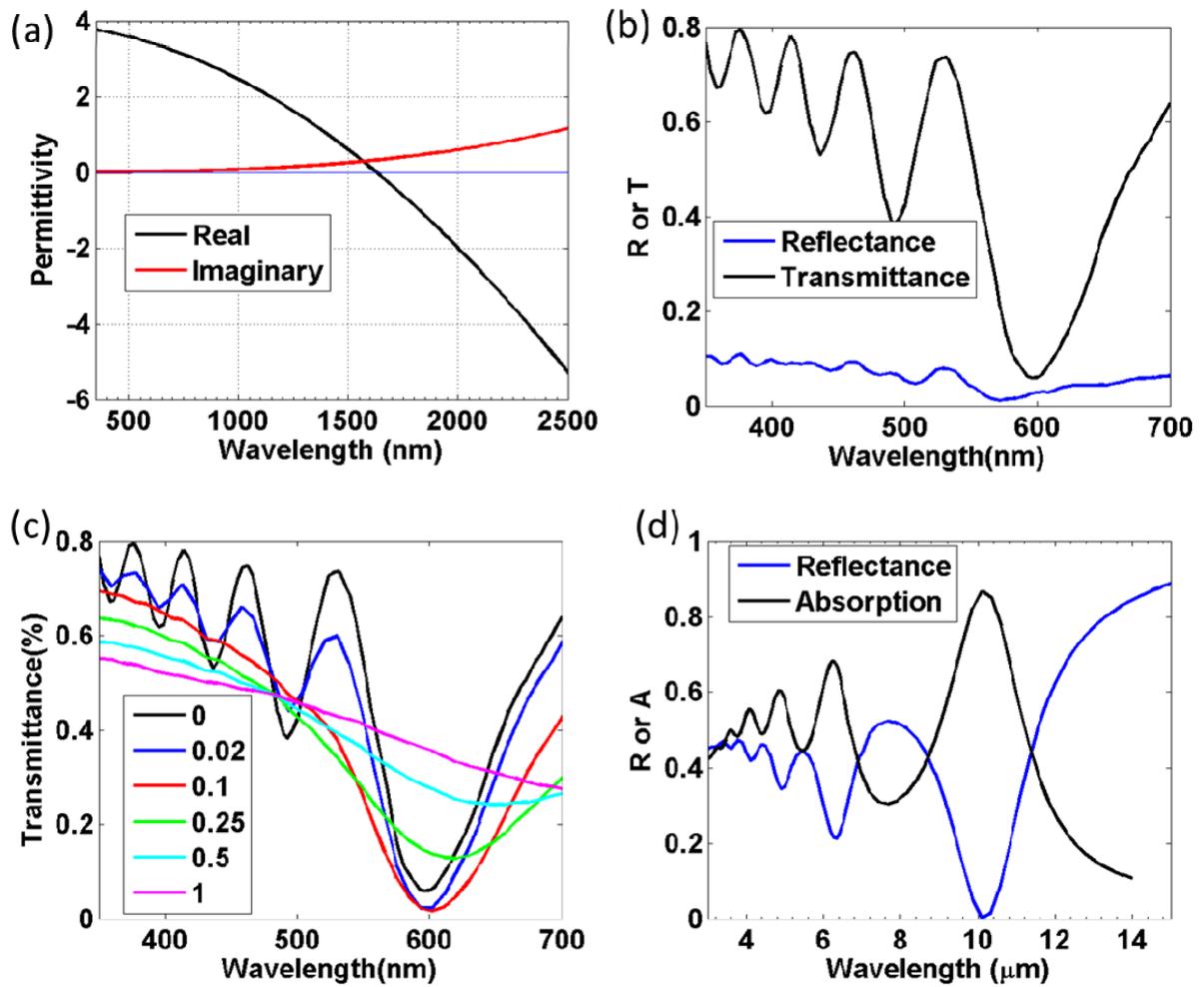

**Figure 2.** (a) Fitted Permittivity of the sample with Drude model. (b) Reflectance and transmittance spectra calculated with FEM simulation. (c) The dependence of transmittance on artificially added absorption to the original absorption coefficient of ITO, which can be expressed as $k = k_0 + x$, with $k_0$ being the original absorption coefficient and $x$ being the artificially added coefficient. $x$ was increased from 0 to 1. (d) The infrared reflectance and absorption calculated from simulation. The polarization of the incident wave is p-polarized, with polar angle equal to 76° and azimuthal angle equal to 0°.

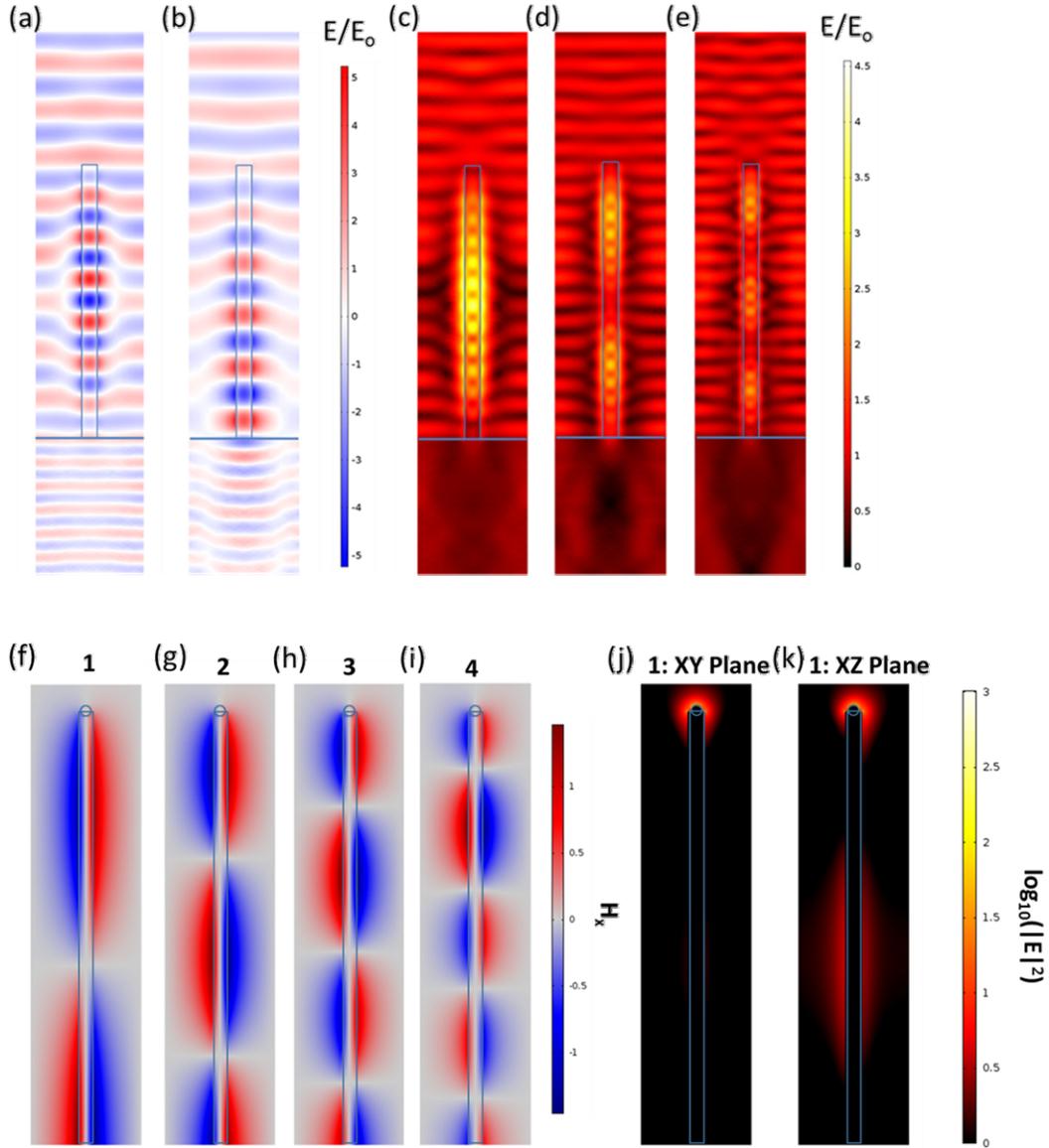

**Figure 3.** (a) and (b) are the electric field $E_x$ normalized to the incident electric field $E_o$ at I and I*. (c)-(e) are the electric field strength of the first three nodes marked in Figure 4 (I, II, and III). All five plots are x-z plane cut at the center of the nanorod. The wave was launch from the top, with electric field polarized along x-axis. The horizontal thick blue line marks the boundary between air (n = 1) and substrate (n = 2). The scattered magnetic field distributions around the nanorod in the array shown in Figure 1(c) have been plotted in (f) to (i) for mode 1 to 4, respectively. The component chosen lies in the scattering plane, orthogonal to the incident light magnetic field direction. The plane cut for (f) to (i) is in the scattering plane (x-z plane). In (j) and (k), the electric field intensity around the same nanorod is plotted for two plane cuts. y-z-Plane is orthogonal to the scattering plan while x-z plane is in the scattering plane. All plane cuts are at the center of the nanorod. The outline of the gold at the top of the nanorod and the nanorod are in blue.

In order to understand the nature of the oscillation, we pick spectral positions I and I* (marked in Figure **2(b)**) for our study. The near field patterns were obtained from simulation as shown in Figure 3**(a)&(b)** respectively. We see that at the top of the nanorods, the electromagnetic wave distribution has a phase front normal to the rod. When the wave propagates in the region with the nanorod, part of the wave was guided in the rod and part of the wave was propagating in the free space. The guided wave propagates with slightly higher phase velocity than the free space wave. It can be seen clearly that at the peak of transmittance, the wave propagated in air and the wave propagated in the nanorods becomes in phase when they reach the boundary between array and substrate. On the other hand, out of phase interference results a transmission minimum as shown in Figure 3**(b)**.

The other oscillations at shorter wavelengths with decreasing magnitude are the higher order interference of the same nature. In Figure 3**(c) to (e)**, the profiles of electric field strength were plotted for the first three transmission maxima. Increasing number of nodes can be seen clearly. Based on the number of nodes, we concluded that I to III are the first to third order resonances. The electric field strength also decreases as shown in the near field plots, which is a result of the decreasing coupling efficiency of wave from free space to nanorod with the decrease of wavelength. The highest field enhancement to the background electric field is around 5 for the first order mode. The large electric field on the surface of the nanorods at this mode is very important for the applications of nanorod array structures to nanostructured solar cells and optoelectronic devices. In most designs, a core-shell structure is used,[8, 34] so the concentration of electric field at the surface can potentially enhance the absorption of light in the active region.

Using the same model, the infrared reflectance spectrum of the 5 μm array was calculated and presented in Figure 2**(d)**, which also matches well with our experimental measurements too. The permittivity at this wavelength regime is metallic, with the real part of the permittivity being a large negative value and imaginary part of the permittivity having large positive value. We expect that the formation of the resonance patterns shown in Figure 2**(c)** to be from the excitation of collective electron motion in the nanorods; i.e., they are plasmon modes, which are well studied in metallic systems like silver and gold nanowires.[35] Due to the damping of electron motion in metal, the absorption is also at the peak value of resonances. Plotted together with the reflectance, we have the absorption curve in Figure 2**(d)**, which indeed peaks at various resonance dips. In order to differentiate the numbering with respect to the modes in the visible region, we have labeled them with Arabic numbers 1 to 4.

As these modes in the infrared are plasmonic, they can be regarded as surface waveguide modes, contrary to the bulk guided modes in the visible regime discussed above. Because the length of the nanorods is comparable to the wavelength and the cross-section is much smaller than the wavelength (d = 185 nm << λ (from 3 μm to 14 μm)), only p-polarized light, which has electric field component along the long axis of the nanorod, can effectively couple into the surface modes. These guided modes (or longitudinal surface plasmons) are intrinsically lossy, which will cause significant absorption in resonances.[36] This is consistent with what we observed experimentally, in which the array is almost invisible from s-polarized light, while strongly interact with p-polarized light.

In Figure 3 we also plotted the simulated magnetic field patterns around the nanorods in response to the incident wave. The x-component of magnetic field in xz-plane was plotted, on x-z-plane, which is the scattering plane. The chosen field component is orthogonal to that of the incident wave (y-component only), so it is purely the scattered field from the light-nanorod-array interaction. As these modes are collective electron motion along the long axis of the nanorod, the magnetic field generated around the nanorod is then proportional to the current strength in the nanorod. In other words, the maxima around the nanorod that we see in Figure 3**(f)** to **(i)** correspond to the nanorod region with largest current. As the tip of the nanorod is the boundary between metallic nanorod and dielectric (air), current is the smallest, gives a node of oscillation, while the bottom boundary is between metallic nanorod and metallic ITO film, so the current is shorted into the film, giving an anti-node of magnetic field oscillation. This is consistent with the mode characteristic of plasmons. The number of nodes can be seen to increase as wavelength decreases. On the other hand, the coupling efficiency decreases with increasing nodes, leading to smaller oscillation strength we see in Figure 2**(c)**.

In Figure 3**(j) and (k)**, we have plotted the electric field intensity distribution around the nanorod at the resonance position of mode 1 with two different cut-planes. It can be clearly seen that the electric field intensity is enhanced and strongest at the top tip of the nanorod, a combination of lightning rod effect and plasmonic resonance.

By replacing the square cross-section of our nanorods with a circular cross-section of the same circumference, the known solutions for $HE_{11}$ mode for the circular dielectric waveguide in the visible and $TM^0$ mode in the infrared are obtained with graphic method.[37] In Figure 4**(a) and (b)**, the results of normalized propagation constant ($\beta_z/\beta_0$) obtained based on the equations are plotted together with the results from numerical mode solver of COMSOL with the actual square cross-section, from which we can conclude that the propagation constant is not sensitive to the shape of the cross-section. The normalized attenuation constants $\alpha/\beta_0$ for both regions are also plotted, from which it is quite convincing that significant absorption occurs for infrared surface waveguide (see Figure 4**(c) and (d)**) but not for visible waveguide modes.

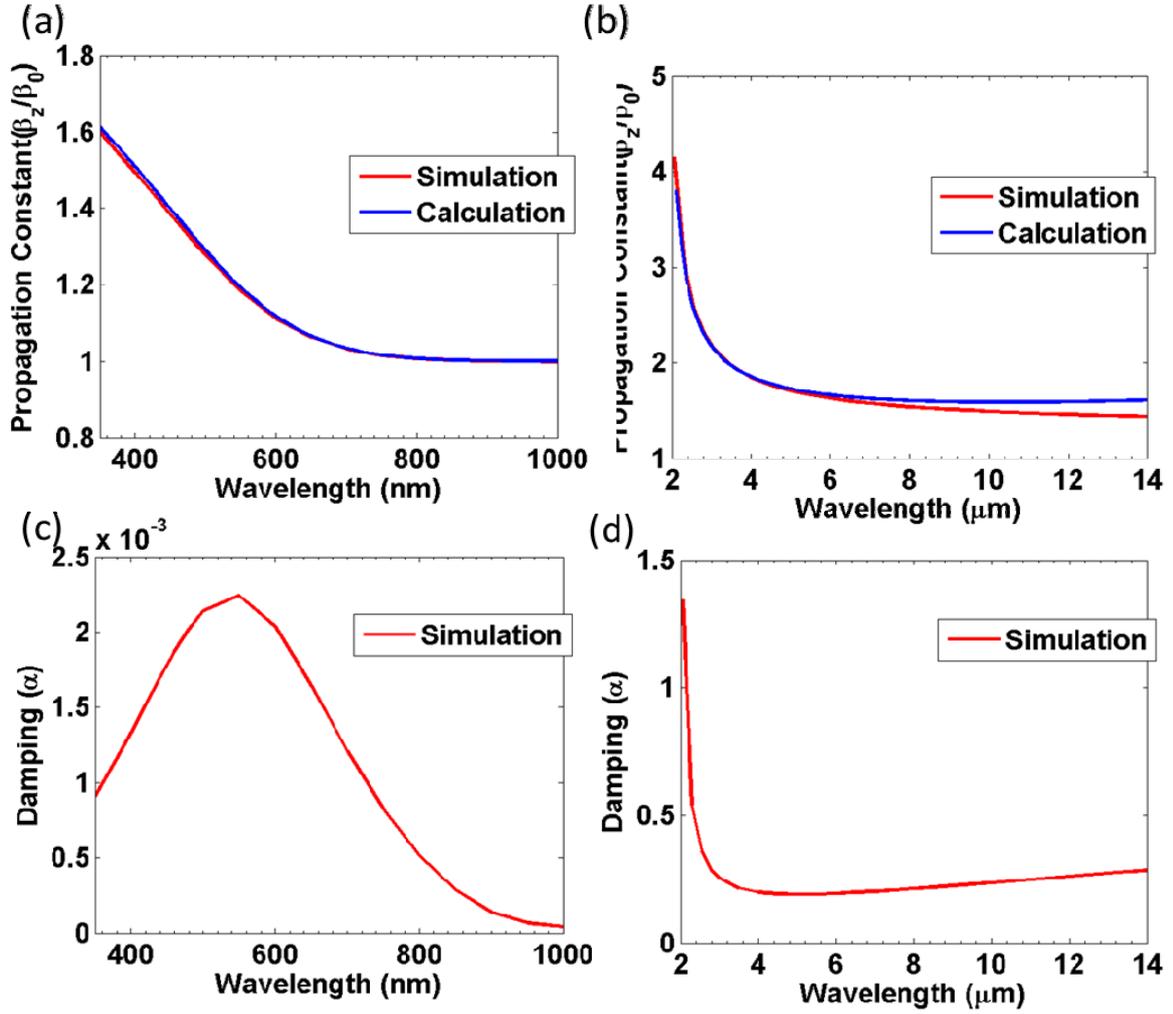

**Figure 4.** Propagation constant calculated with circular cross-section matches well with simulation for infrared region (a) and visible region (b). (c) and (d) are plots of attenuation constant α calculated for both regions of interest.

Based on the propagation constant, we can proceed to the calculation of resonance positions observed in the experiments. As discussed above, in the visible region, the minima are resulted from the out-of-phase interference of guided wave with free-space propagating wave. As light interacts with the nanorod array, the wave guided by the nanorods experiences different phase accumulation comparing to the wave traveled in between the nanorods. At the end of interaction, if the accumulated phase difference is $(2n\text{-}1)\,\pi$, the transmission will be at its minimum as the free-space wave is out of phase with the guided wave through the nanorods. We designate the order of the modes as *n*, which will give a minimum when it is an integer. The general formula for different order *n* is,

$$\lambda_n = \left(\frac{\beta_z}{\beta_0} - 1\right)\frac{2h}{2n-1}, \quad (2)$$

The result calculated based on **Equation (2)** are plotted together with the spectral positions of experimentally measured transmission minima in Figure 5**(a)**. The agreement of this calculation with experimental results confirms that our postulation.

In the infrared regime, the absorption maxima observed are caused by the standing wave formed in the nanorods, just like the radio frequency (RF) antennas, with a correction factor rising from the effective wavelength of the waveguide on antennas are different from free space wavelength.[38] This effective wavelength scaling was taken care with the propagation constant calculated (shown in Figure 4**(b)**). Without considering the end effect, we can subsequently calculate the standing-wave (antenna) mode numbers from the propagation constant using the following equation,

$$\lambda_n = \frac{2h\beta_z}{n\beta_0}, \quad (3)$$

, where *h* is the height of the nanorod, $\lambda_n$ is the resonant wavelength. Integers of *n* are the orders of resonances. For example, first order resonance occurs at *n* =1. We have plotted the curve based on **Equation (3)** and compare it with the modes observed from the experiments in Figure 1**(d)**. It is noted that the experimental modes are observed at a shorter wavelength than predicted from **Equation (3)**. There are two reasons. Firstly, the end effect was not considered. The two ends of the nanorod are boundaries with impedance mismatches, which are determined by both the geometry and the dielectric contrast. Secondly, the coupling among the antennas is not negligible. This coupling is the mutual induction caused by the magnetic field generated by the current in the surrounding antennas. The coupling strengthens the restoring potential of the electrons, therefore, blue-shifts the resonant frequency with increasing coupling magnitude, which agrees with our observation of modes are occurring at shorter wavelength than predicted by standing wave calculation. For higher order resonances, both end effect and coupling are less significant, thus better agreement was observed.

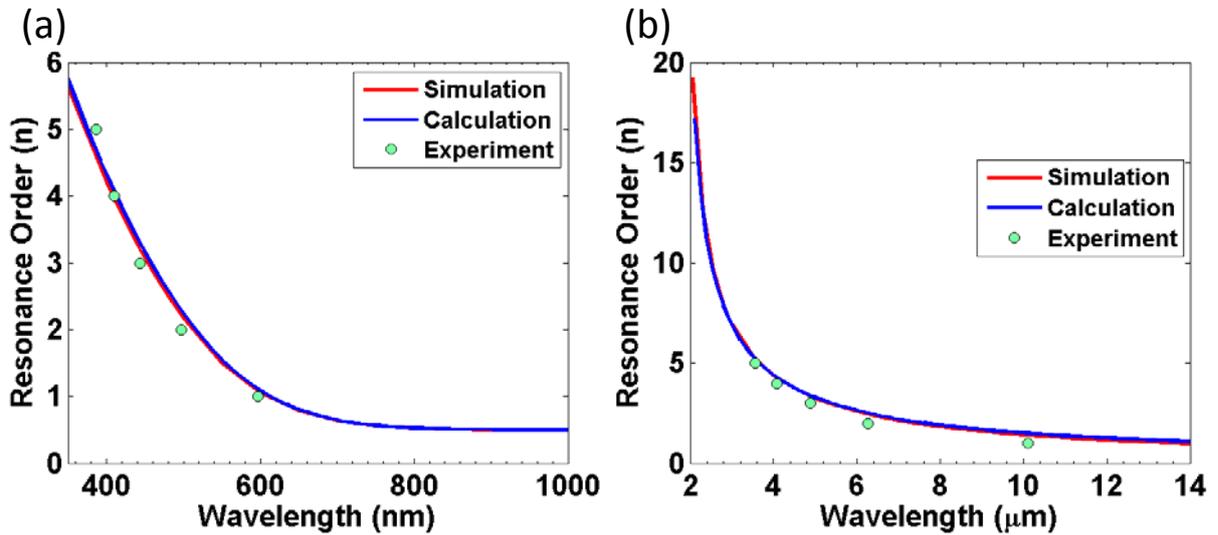

**Figure 1.** Mode number calculated and its comparison with experimental data based on equation (2) for the visible region as shown in (a) and based on equation 3 for the infrared region as shown in (b).

## Conclusions

In conclusion, we analyzed the resonances observed in ordered ITO nanorod array with uniform diameter and height using a unified waveguide approach. Excellent agreements were found among experiment, simulation, and theoretical analysis. A very surprising finding is that the resonances observed in two spectral regions have quite different mechanisms. It would normally be regarded that the resonances in the dielectric waveguide is a result of cavity in the nanorods; however, it is verified to be the interference between free-space wave and guided wave with ITO nanorod array. On the other hand, the waveguide modes in the infrared are standing waves on the surface of nanorods. The conclusions from our study differ from what has been reported previously for semiconductor nanowires, where the resonances in the visible region were considered as standing-wave resonance entirely due to the guided wave in nanorods.[33, 39-42] The geometric cross-section of nanorod array only occupies 2.4% the space but resulted in less than 10% transmission at the first order destructive interference, which implies a strong coupling of light with the nanorod array. In both types of resonances, the fields are greatly enhanced at the vicinity of the surfaces of nanorods, which can be potentially applied to next generation wavelength-selective photodetector, photovoltaics, modulators, and other optoelectronic devices. From the simple analytical equations presented here, the resonant wavelengths can be calculated easily, which are very useful for designing the nanorod-array structure with desired properties.

**Acknowledgement**

Supported by NSF funding (DMR-1121262 and DMR-0843962), QUEST computational resources (Project p20194 and Project p20447) and Center for Nanoscale Materials in Argonne National Laboratory (Project CNM 30831 and Porject CNM 36120). Various characterizations were done in NUANCE center and KECK II Facilities in Northwestern University. The NUANCE Center and KECK II Facilities are supported by the NSF-NSEC, NSF-MRSEC, Keck Foundation, the State of Illinois, and Northwestern University. E-beam lithography was performed with JEOL-9300 in the Center for Nanoscale materials at Argonne National Laboratory. Use of the Center for Nanoscale Materials was supported by the U. S. Department of Energy, Office of Science, Office of Basic Energy Sciences, under Contract No. DE-AC02-06CH11357.